\documentclass[12pt,preprint]{aastex}
\def\gs{\mathrel{\raise0.35ex\hbox{$\scriptstyle >$}\kern-0.6em \lower0.40ex\hbo
x{{$\scriptstyle \sim$}}}}
\def\ls{\mathrel{\raise0.35ex\hbox{$\scriptstyle <$}\kern-0.6em \lower0.40ex\hbo
x{{$\scriptstyle \sim$}}}}

\received{1st July 2001}
\accepted{16th July 2001}

\shorttitle{Spectroscopic observation of galaxies}
\shortauthors{Mobasher et al.}

\begin{document}
 
\title{A photometric and spectroscopic study of dwarf and giant galaxies 
in the Coma cluster -  
II. Spectroscopic observations
\footnote{Based on observations made with the William Herschel Telescope 
operated on the island of La Palma by the Isaac Newton Group in the 
Spanish Observatorio del Roque de los Muchachos of the Instituto de 
Astrofisica de Canarias.}}
\author{
Bahram\ Mobasher,$^{\!}$\altaffilmark{2,3}
Terry J.\ Bridges,$^{\!}$\altaffilmark{4}
Dave Carter,$^{\!}$\altaffilmark{5}
Bianca M. Poggianti,$^{\!}$\altaffilmark{6}
Y. Komiyama,$^{\!}$\altaffilmark{7}
N. Kashikawa,$^{\!}$\altaffilmark{8}
M. Doi,$^{\!}$\altaffilmark{9} 
M. Iye,$^{\!}$\altaffilmark{8}
S. Okamura,$^{\!}$\altaffilmark{10,11} 
M. Sekiguchi,$^{\!}$\altaffilmark{12}
K. Shimasaku,$^{\!}$\altaffilmark{11}  
M. Yagi,$^{\!}$\altaffilmark{8}
N. Yasuda$^{\!}$\altaffilmark{8}
}
 
\smallskip

\affil{\scriptsize 2) Space Telescope Science Institute, 3700 San Martin Drive, Baltimore, MD 21218, USA}
\affil{\scriptsize 3) Affiliated with the Space Sciences Department of the 
European Space Agency}
\affil{\scriptsize 4) Anglo-Australian Observatory, PO Box 296, Epping, NSW 2121, Australia}
\affil{\scriptsize 5)  Astrophysics Research Institute, Liverpool John Moores University, Twelve Quays House, Egerton Wharf, Birkenhead, Wirral, CH41 1LD, UK}
\affil{\scriptsize 6) Osservatorio Astronomico di Padova, vicolo dell'Osservatorio 5, 35122 Padova, Italy}
\affil{\scriptsize 7) Subaru Telescope, 650 North Aohoku Place, Hilo, HI 96720, USA}
\affil{\scriptsize 8) National Astronomical Observatory, Mitaka, Tokyo, 181-8588, Japan}
\affil{\scriptsize 9) Institute of Astronomy, School of Science, University of
Tokyo, Mitaka, 181-0015, Japan}
\affil{\scriptsize 10) Research Center for the Early Universe, School of
Science, University of Tokyo, Tokyo 113-0033, Japan}
\affil{\scriptsize 11) Department of Astronomy, University of Tokyo,
Bunkyo-ku, Tokyo 113-0033, Japan}
\affil{\scriptsize 12) Institute for Cosmic Ray Research, University of Tokyo,
Kashiwa, Chiba 277-8582, Japan}

\begin{abstract}

This is the second paper in a series studying the photometric and spectroscopic
properties of galaxies of different luminosities in the Coma cluster. 
We present the sample selection, spectroscopic observations 
and completeness functions. To study the spectral properties of galaxies 
as a function of their local environment, two fields were selected for 
spectroscopic observations to cover both the core (Coma1) and
outskirts (ie. south-west of the core and centered on NGC4839)- (Coma3) 
of the cluster. To maximise the efficiency of
spectroscopic observations, two sub-samples were defined, consisting
of `bright' and `faint' galaxies, both drawn from magnitude-limited
parent samples. Medium resolution spectroscopy (6-9 \AA) was then 
carried out for a total of 490 galaxies in both fields
(302 in Coma1 and 188 in Coma3), using the WYFFOS multi-fiber 
spectrograph on the William Herschel Telescope. The galaxies
cover a range of $12 < R < 20$, corresponding to $-23 < M_R < -15$ (H$_0=65$
km/sec/Mpc.  The redshifts for these galaxies 
are measured with an accuracy of 100 km/sec. 
The spectral line strengths and equivalent widths  
are also measured for the same galaxies and analysed in
Poggianti et al (2001- paper III). A total of 189 (Coma1) and 90 (Coma3)
galaxies are identified as members of the Coma cluster. An analysis of
the colors show that only two members of the Coma cluster in our sample have
$B-R > 2$. The completeness functions for our sample are calculated. These
show that the `bright sample' is $65\%$ complete at $R < 17$ mag, becoming
increasingly incomplete towards fainter magnitudes, while the `faint sample'  
follows a monotonically decreasing completeness function at $R > 19$ mag.

\end{abstract}


\section{Introduction}

In this series of papers we present the results from a wide-area photometric 
and spectroscopic survey of galaxies at the core and outskirts of 
the Coma cluster. In the first paper details of the photometric observations
and data reduction are discussed (Komiyama et al 2001; paper I). 
This is the second paper in the series, presenting the
spectroscopic sample selection, observations and completeness
functions. The third paper performs an analysis of the diagnostic line
indices for the entire spectroscopic sample (Poggianti et al 2001a). 
Using the spectroscopic data here, a study of the ages of elliptical and S0 
galaxies is also carried out (Poggianti et al 2001b).  
The future papers in this series present spectroscopic 
comparison with galaxies in intermediate redshift clusters, 
study of the color-magnitude relation at the core and outskirts of the Coma
cluster and the source of its scatter, luminosity function of
galaxies in the Coma and its dependence on the local environment
and a study of the spatial distribution and dynamics of galaxies 
in this cluster.  

Clusters of galaxies provide ideal environments for studying formation
and evolution of galaxies.  By 
studying galaxies located at the center and outer regions of a single 
cluster, one 
expects to minimise the parameter space, affecting their evolutionary
properties, only to those of the local density. This also provides a 
continous sequence for study of galaxy evolution from dense (cores
of clusters) to less dense (outskirts of clusters) and general field. 
However, due to technological limitations, caused by small format CCDs, 
such studies have so far been concentrated only to 
distant clusters ($z> 0.4$), where only the bright-end of the luminosity 
function could be sampled (Dressler et al. 1999; van Dokkum et al 1998). 
For these clusters, one also has the addtional problem of 
obtaining medium and high resolution spectroscopic data for a
statistically significant number of galaxies over a wide range in 
luminosity. Moreover, it is not clear if the evolutionary history of 
galaxies in the intermediate 
to high redshift clusters are indeed similar to those observed in the 
local Universe, 
indicating the need for a detailed study of the physical properties
of galaxies at the core and outskirts of nearby clusters. In recent years,  
such investigation has become possible due to
the advent of mosaic CCDs and multi-object fiber spectrographs, allowing
photometric and spectroscopic observations over large solid angles.

Given the above requirements, we started a large photometric and spectroscopic
survey of galaxies at both the core and outskirts of the Coma cluster. 
The aim of this is to perform a detailed study of the
photometric (luminosity, surface brightness, color) and spectroscopic 
(indices sensitive to metallicity and star-formation) parameters 
for galaxies at different radial distances from the center of
a rich local cluster (ie. the Coma). There are two main differences between
the present and previous such studies. Firstly, this covers a significantly
larger area than previous spectroscopic surveys of the Coma cluster, 
providing homogeneous dataset for a large sample at different environments. 
Secondly, it
extends the medium resolution spectroscopy of galaxies to faint
luminosities ($R\sim 19-20$), allowing comparison of 
 properties of galaxies in the range $12 < R < 20$ (corresponding to
$-23 < M_R < -15 $ for H$_0 = 65$ Km/sec/Mpc). 

A summary of the photometric observations followed by spectroscopic
sample selection and spectroscopic observations are presented in section 2.
Section 3 presents the selection function for the spectroscopic sample
in this study. This is followed in section 4 by a summary of the main
results. 

\section{Observations}

\subsection{Photometry}

The photometric observations and data reduction are explained in detail in 
paper I. Briefly, a total area of 2.22 deg.$^2$ was surveyed, with a 
sub-area of 1.375 deg$^2$ covered in both the B and R bands. This consists of
five contiguous fields at the core and outskirts of the Coma cluster.  
The observations
were carried out in the Johnson B and Cousins R bands and are complete 
to $R=21$ mag ($B=22.5$ mag.). To allow for a statistical estimate of 
the background contamination, 
a control field (SA57) close to the Coma cluster was also observed under
similar conditions and to the same depth. 

Source detection was carried out using the `connected pixel' algorithm. 
An object is entered into the catalogue if it extends over a number of 
connected pixels larger than a threshold value (taken to be 30), 
with all the pixels having values above the detection threshold
(1.5 $\sigma_{sky}$). Detailed discussion of source detection and
star/galaxy separation is given in paper I. 
Photometric catalogues were then constructed for each field, consisting of 
the magnitude (isophotal, aperture, Petrosian), 
surface brightness (mean, effective), radius (effective, Petrosian, Kron)
and the compactness parameter. In this paper, we use the magnitudes 
measured over a circular aperture of diameter 3 times the Kron radius, 
$r_{k}$. This is an intensity-weighted radius defined as 
$$r_k = {\sum_i (r_i\times I_i)\over \sum_i I_i} $$
where $r_i$ and $I_i$ are respectively, 
the radius and intensity of the $i$th pixel, 
with the summations over all pixels above the detection threshold. 
This has the advantage
of scaling magnitudes proportional to the size of the galaxies
(using larger apertures for larger galaxies)
and hence, measuring the light from all of the galaxy (bulge+disk). 
$B-R$ colors are also determined over an aperture of diameter 
$3\times r_{k}$ 
and agree within $2\%$ with those measured over a fixed aperture of 10 arcsec.
diameter. Both magnitudes and colors are close to the total values. 
The estimated photometric accuracy of the present data is 0.03 mag
(R-band), 0.04 mag (B-band) and 0.05 mag ($B-R$) as estimated from 
repeated observations of galaxies which lie in the overlapping regions
between the contiguous fields. 
The astrometry, performed with the APM, is accurate
to better than 0.5 arcsec, sufficient for the fiber spectroscopy
discussed in the next section.

Two fields (each $32.5 \times 50.8$ arcmin) were
selected for medium resolution fiber spectroscopy. These were
chosen to cover areas with large density contrasts, consisting of
the core field at the cluster center (Coma1) and a field south-west of
the cluster, centered on NGC4839 and the X-ray secondary peak (Coma3). 
The coordinates of the centers of the two 
spectroscopic fields and the total number of galaxies detected in each
field to the magnitude limit of the sample 
are listed in Table 1. The criteria used for 
selecting the sample for spectroscopic observations are discussed in
the next section.

\begin{table*}
\pagestyle{empty}
\caption{Coordinates of the centers of the two spectroscopic fields in Coma}
\begin{tabular}{lccc}
   &            &            &     \\
   &  RA(J2000) & Dec(J2000) &  $n$ \\
   &            &            &  \\
Coma1 (center field) &  12h\ 59m\ 45.17s & 27$^\circ$\ 57$'$\ 53.1$''$\ & 2744 \\
Coma3 (NGC4839 field)&  12h\ 57m\ 28.48s & 27$^\circ$\ 08$'$\ 08.5$''$\ & 2210 \\

\end{tabular}
\end{table*}

\subsection{Selection criteria for the spectroscopic sample} 

The aims of the spectroscopic observations are to identify members
of the cluster in the Coma1 and Coma3 fields and to measure their spectral line 
indices. The latter requires medium resolution spectroscopy of galaxies
which are confirmed members of the Coma cluster.  
In order to maximise the efficiency of spectroscopic observations, 
different criteria were used for selecting galaxies by dividing the 
photometric sample into two broad classes. First:  
galaxies which already have redshifts from other studies and are 
spectroscopically 
confirmed members of the Coma cluster. These are mostly high luminosity
galaxies to which we refer as the `bright sample'. Second: objects with
no available redshifts which are, on average, fainter than the first group. 
This is referred to as the `faint sample'. Therefore, the aim of the 
medium resolution
fiber spectroscopy is to measure the spectroscopic line indices for 
galaxies in the bright sample (which all are confirmed cluster members) 
and determine both redshifts and line indices for galaxies in the 
faint sample. 
Different selection criteria used for the two samples are explained in
detail below.

\noindent a). The bright spectroscopic sample:   
This sample was taken from the compilation in Colless \& Dunn (1996),
and an unpublished list of redshifts for Coma galaxies kindly provided 
by M. Colless. As this compilation is rather inhomogeneous, using objects
with different selection criteria, 
it is important to investigate possible sources of bias which could be 
reflected in our spectroscopic sample. 

The catalogue by Colless \& Dunn (1996), consisting of 552 redshifts, 
is selected from the photographic plate survey by Godwin, Metcalfe 
and Peach (1983), which is a magnitude limited sample. This includes
redshifts measured by these authors and those compiled 
from Kent \& Gunn (1982), van Haarlem et al (1993)
and Caldwell et al (1993). We checked these samples and apart from
Caldwell et al, which mainly contains early-type galaxies, did not find 
evidence for any morphological bias. As the Caldwell et al.
sample only constitutes a small fraction of the present survey ($2\%$), 
this is not expected to introduce a bias in the final spectroscopic 
catalogue. However, it is possible that the spectroscopic sample is slightly 
biased against late-type galaxies which, on average, have a fainter 
characteristic luminosity than the early-type systems. Apart from this, 
there is no known bias in terms of morphological type or color in 
this sample. 
Therefore, we consider this a magnitude limited sample with no significant 
bias.

From the compiled sample, a total of 257 and 123 galaxies were found to be
in the Coma1 and Coma3 fields respectively. All the galaxies in this 
sample have $R < 18$ mag, with their R-band magnitude and
$B-R$ color distributions presented in Figure 1. 
The candidates for spectroscopic observations were then selected to have a 
uniform distribution over the magnitude range covered by their parent sample. 
These consist of 138 and 85 galaxies in Coma1 and
Coma3, respectively. The magnitude and color distributions for these
galaxies, also presented in Figure 1, show a similar distribution as
their parent sample.

\begin{figure}
\plotone{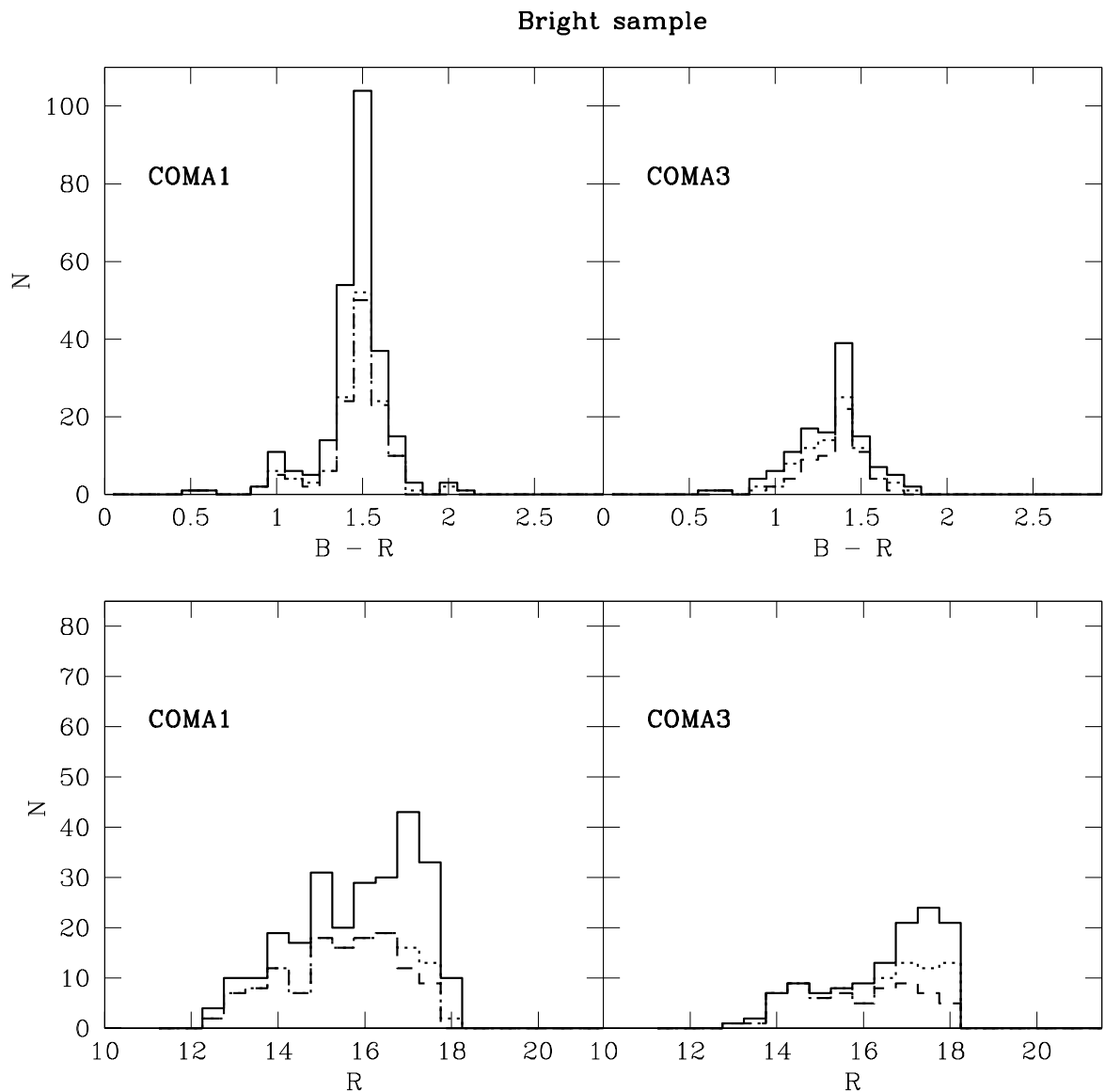}
\caption{ R-band magnitude and $B-R$ color distributions for 
galaxies in the bright spectroscopic sample 
in the Coma1 and Coma3 fields. The histograms correspond to
the parent sample from which the spectroscopic candidates were selected
(all confirmed members of the Coma cluster)-
(solid line); the objects drawn from the parent sample for
medium resolution spectroscopic observations (dotted line); objects for
which redshifts and spectral line strengths were obtained (dashed line).\label{fig1}}
\end{figure}

In order to make the maximum use of the fiber configurations, a few spare
fibers were assigned to objects with 
$R < 18$ mag. and no available redshifts. These objects, which are
selected randomly, were included
where no source with spectroscopically confirmed membership was available.

\noindent b). The faint spectroscopic sample:   
The `faint sample' for spectroscopic observations was selected from
the photometric survey described in paper I, 
based on their colors and magnitudes. 
The $(B-R)-R$ color-magnitude diagram for galaxies in 
the entire photometric survey of Coma1 and Coma3 fields 
is presented in Figure 2. The objects in the faint sample are selected
to satisfy $ 17.5 < R < 20$ (corresponding to $ -17.5 < M_R < -15$) 
and $1 < B-R < 2$. The area covered by
these galaxies in the color-magnitude diagram is shown in Figure 2 
by a rectangle. 
The brighter magnitude limit is adopted to allow dwarf galaxies
into the sample 
while the color boundaries eliminate a large fraction of non-cluster galaxies
(Secker 1996).  
The slight overlap in magnitude between the `bright' and `faint' samples 
allows inclusion into the spectroscopic sample of some relatively bright 
galaxies ($R\sim 17.5$) for which redshifts are not available. 
The faint magnitude limit is set by the completeness requirement and 
feasibility in getting the required S/N in a reasonable exposure time. 
To explore 
the nature of galaxies satisfying this requirement and the effect of
our color selection criteria, we compute the spectral energy distributions and
colors for models with a broad range of star-formation histories, mimicking
ellipticals, spirals (of types Sa, Sb, Sc and Sd), and galaxies
with a current starburst at $z=0.023$ (the redshift of the Coma cluster)-
(Barbaro \& Poggianti 1997). The metallicity evolution was taken into
account, so that the models of the later types have a lower average 
metallicity than those of the earlier types. All but the starburst models
are found to have colors redder than our blue cut-off ($B-R > 1$). Therefore, 
the color boundaries adopted for the faint sample here would {\it only} 
exclude the
starbursts (if present). Figure 2 shows that there are 22 galaxies
with $B-R < 1$ and $17.5 < R < 20$, constituting only 5\% of the total 
spectroscopic sample
(this is the fraction which one would miss by imposing the $B-R$ color 
criteria adopted here). 

\begin{figure}
\plotone{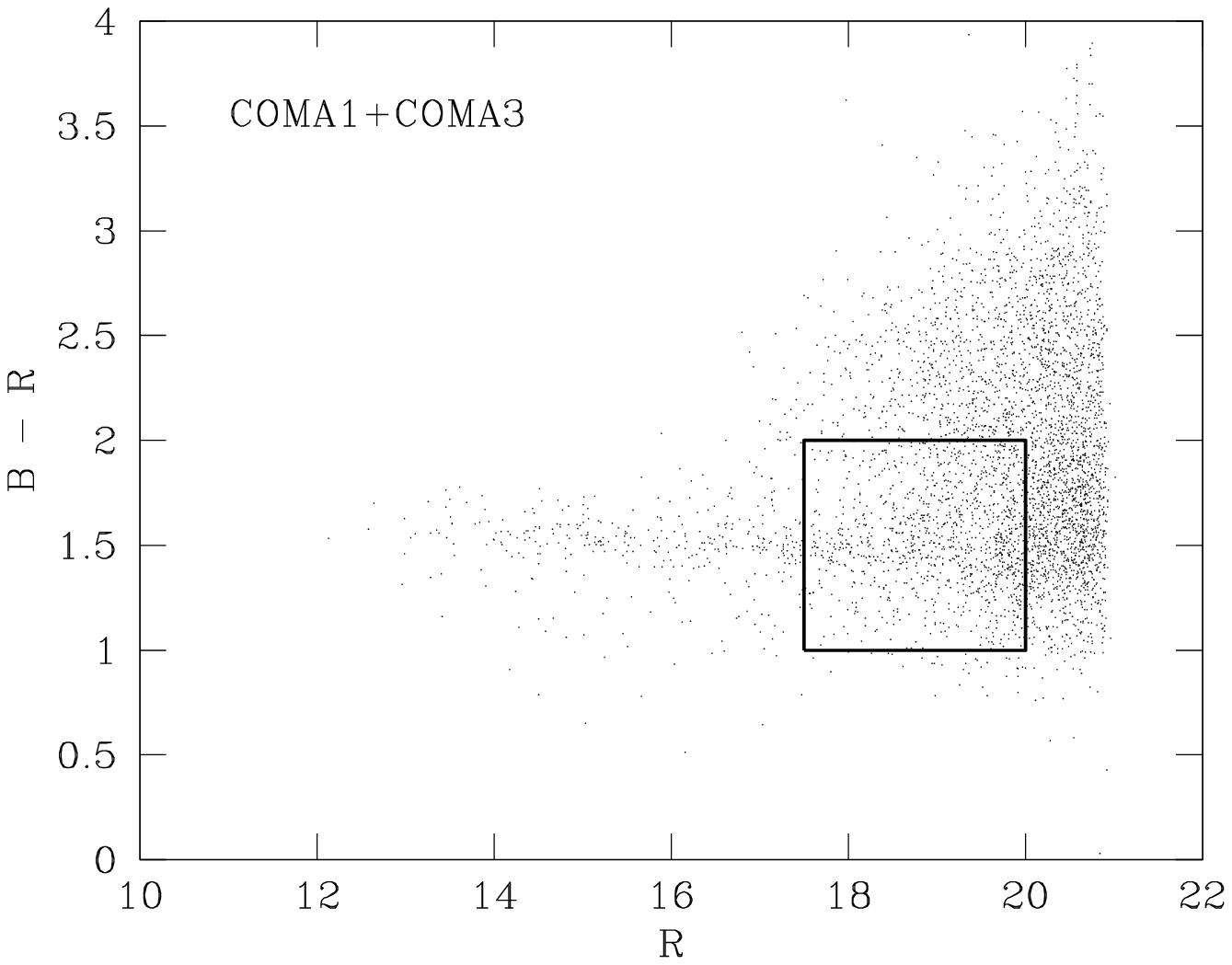}
\caption{  R vs. B-R color-magnitude diagram for the {\it full}
photometric survey in the Coma1 and Coma3 fields. The rectangle
represents the expected domain for members of the Coma cluster 
in the faint sample (see text for details).\label{fig2}}
\end{figure}

A total of 222 and 190 galaxies in Coma1 and Coma3 
satisfy these criteria. The R-band magnitude and B-R color 
distributions for galaxies in this sample are
shown in Figure 3. The spectroscopic candidates were selected randomly from 
this parent sample, consisting of a total
of 83 and 79 galaxies in Coma1 and Coma3 fields respectively.  The
restricting factor in selecting the spectroscopic candidates from the
parent sample here was the total number of fibers which could be allocated
to the objects in a given configuration and the longer exposure time 
required for galaxies in the faint sample. The magnitude
and color distributions for these galaxies are also presented in Figure 3,
where they show a similar distribution to that of their parent sample. 

To further test the selection criteria for the faint sample here, the
color condition was relaxed to explore if any Coma cluster member could be 
excluded from the spectroscopic sample due to its color. This has the effect
of allowing objects with $B-R < 1 $ and $ B-R > 2$ into the sample and 
exploring if such
galaxies could indeed be members of the Coma cluster. This is done by
first removing the $B-R$ color constraint in the selection of galaxies 
in the faint
sample, and then randomly choosing the candidates for spectroscopic
observations. 
The magnitude and color distributions for galaxies 
in this sub-sample are presented in Figure 4 for Coma1 (81 galaxies) 
and Coma3 (24 galaxies) fields.  
This constitutes all the galaxies for which 
reliable spectroscopic data (redshifts and spectral line strengths) 
were measured. The distribution of
galaxies in this sample which are spectroscopically confirmed members of the
Coma cluster (see section 2.3) are also presented in Figure 4 and show that
there are {\it only} two members of the Coma cluster (67822 and 102739) 
with $B-R > 2$ in our sample. 
These are galaxies 67822 and 102739, with $B-R=$ 2.139 and 
2.422
respectively. However, the $B-R$ colors for these objects over 10 arcsec 
diameter apertures are 1.638 (for 67822) and 1.704 (for 102739).
These are both substantially bluer than the $B-R$ colors used here, which
are measured over an aperture of diameter 3 times the Kron radius, 
indicating strong color gradients in these objects.
This confirms that the adopted color criteria here do not bias against
galaxies which are members of the Coma cluster.

\begin{figure}
\plotone{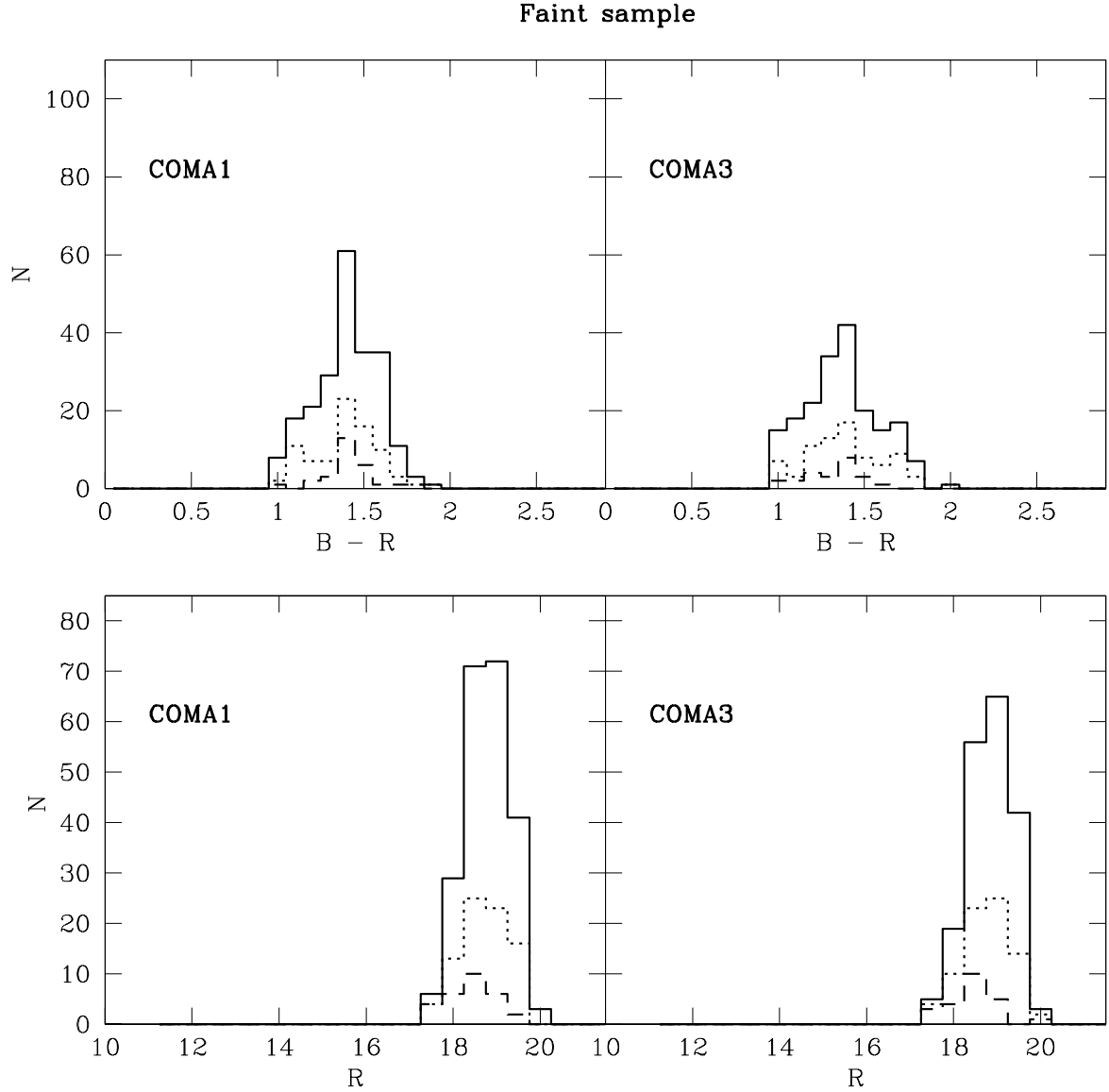}
\caption{ R-band magnitude and $B-R$ color distributions for
the faint sample in Coma1 and Coma3 fields. The histograms correspond to the
parent sample (selected as discussed in the text)-(solid line); objects
randomly selected from the parent sample and for which spectroscopic 
observations were obtained (dotted line); spectroscopically confirmed
members of the Coma cluster (dashed line). \label{fig3}}
\end{figure}

\begin{figure}
\plotone{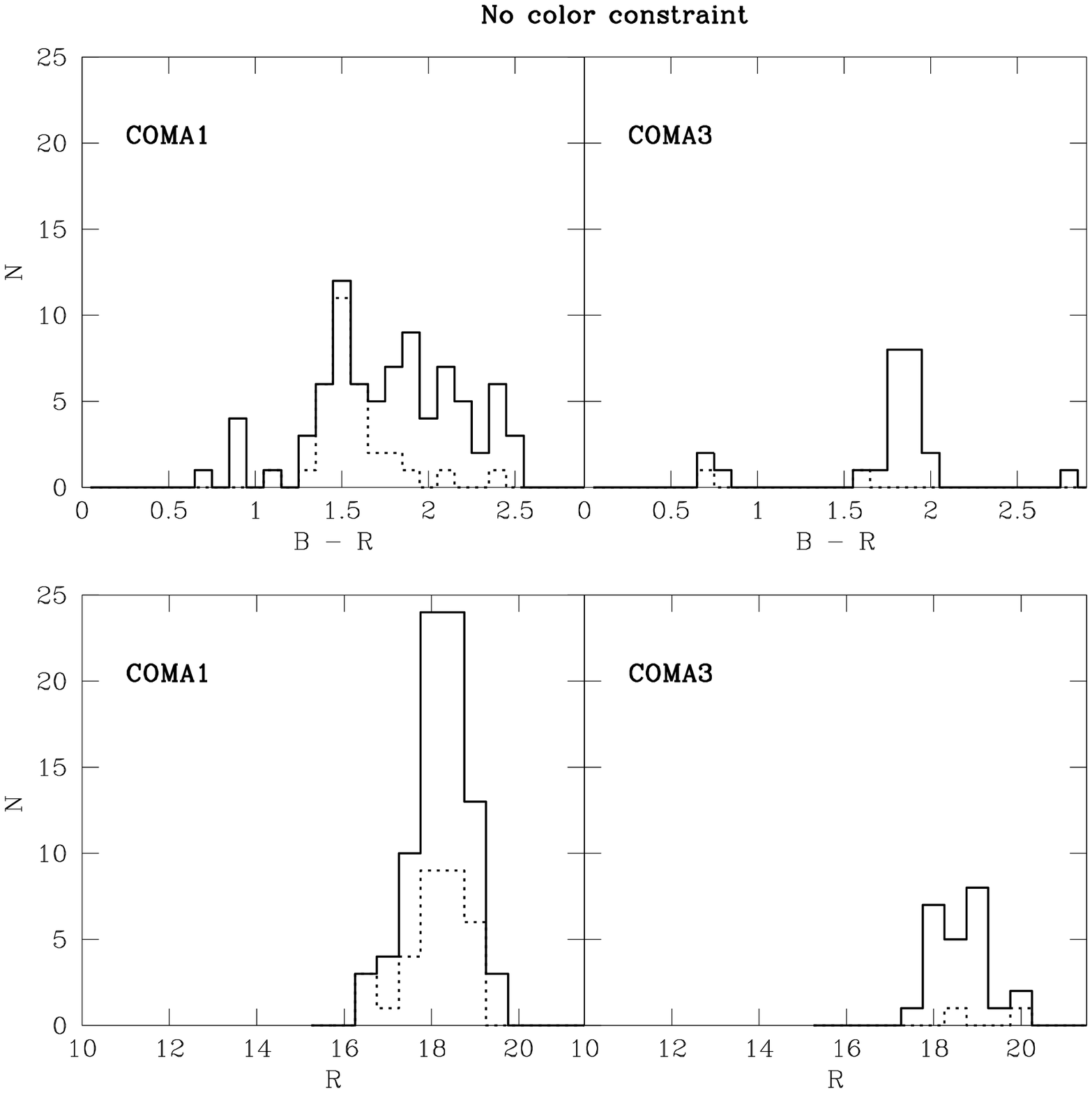}
\caption{ R-band magnitude and $B-R$ color distributions for galaxies from 
the sample for which the color constraint was removed. 
Objects in both the Coma1 and Coma3 fields are shown. 
The solid line corresponds to the distribution for the entire sample
for which spectroscopic data were obtained. The dotted line corresponds to
objects which are spectroscopically confirmed members of
the Coma cluster.\label{fig4}}
\end{figure}

The total number of galaxies in the spectroscopic sample in Coma1 and Coma3,
and those identified as cluster members (Section 2.4) are listed in 
Table 2. The spatial distribution of galaxies in all three 
spectroscopic sub-samples
discussed above are presented for both the Coma1 and Coma3 fields in 
Figure 5. 

\begin{figure}
\plotone{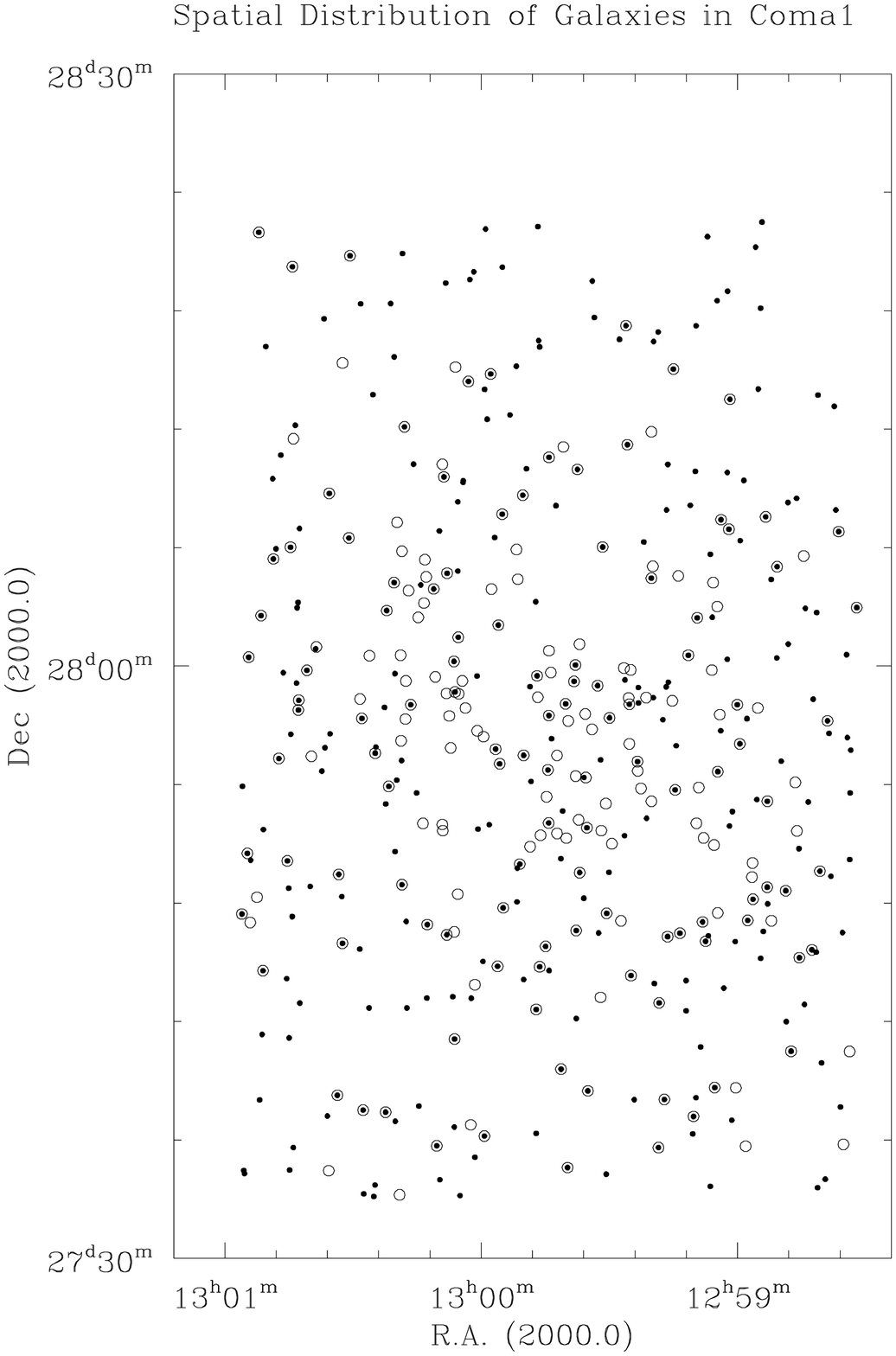}
\end{figure}
\begin{figure}
\plotone{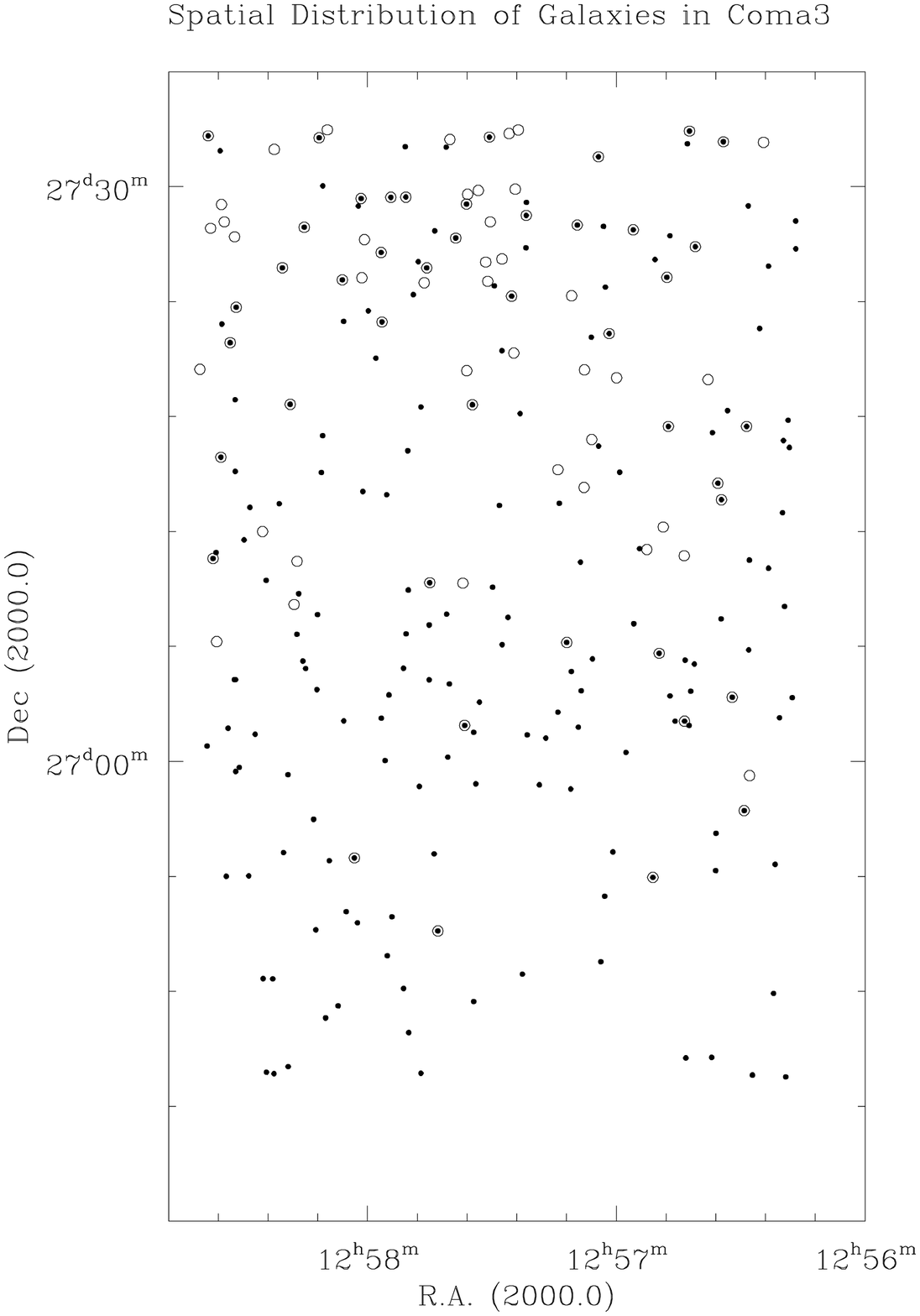}
\caption{Spatial distribution of galaxies 
for which medium resolution spectroscopy was obtained.
This includes objects with spectroscopic data from
the present study (filled circles), objects with available spectroscopic 
redshifts
from Colless \& Dunn (open circles), and objects in common between the two
studies (circle+dot).\label{fig5}}
\end{figure}

The reason for dividing the spectroscopic sample in this study into two
bright and faint samples was two-fold. Firstly, this increases the
efficiency of the spectroscopic observations by allowing different exposure 
times for the two sub-samples in fiber-spectroscopic observations. 
Secondly, this allows separation of the spectroscopic  sample into 
giant (bright) and dwarf (faint) sub-samples. A study of the
spectroscopic properties of dwarf galaxies compared to giants will then be
possible. However, 
the criteria used to define the bright and faint samples here are
independent from galaxy surface brightness. Also, the technique
used to identify the dwarf and giant galaxies in paper III is only based
on their magnitudes, with no surface brightness-based classification.  
As dwarfs have, on average, 
a lower surface brightness than the giants, one expects this to provide an
independent test of the above classification schemes. 
The effective surface brightness for galaxies in this sample are
measured in paper I by assuming the Kron magnitude to correspond to
the total magnitude of galaxies and hence, estimating the effective
radius (the radius containing half the total light of galaxy) for
each galaxy. The mean surface brightness inside the effective radius
of a galaxy is then measured and used as the effective surface brightness. 
The R-band effective  
surface brightness distributions for galaxies in the bright and faint 
spectroscopic samples are compared in Figure 6. This confirms that 
the bright sample has a fairly wide range in effective surface brightness
($19 < \mu_R^e < 23.5 $), implying
that there is no strong surface brightness related bias in this sample. 
However, for the faint sample, the relatively narrower range covered
in surface brightness is likely caused by a bias due to the magnitude
limit of this sample, which excludes galaxies with $\mu_R^e > 24$
mag./arcsec$^2$. 

\begin{figure}
\plotone{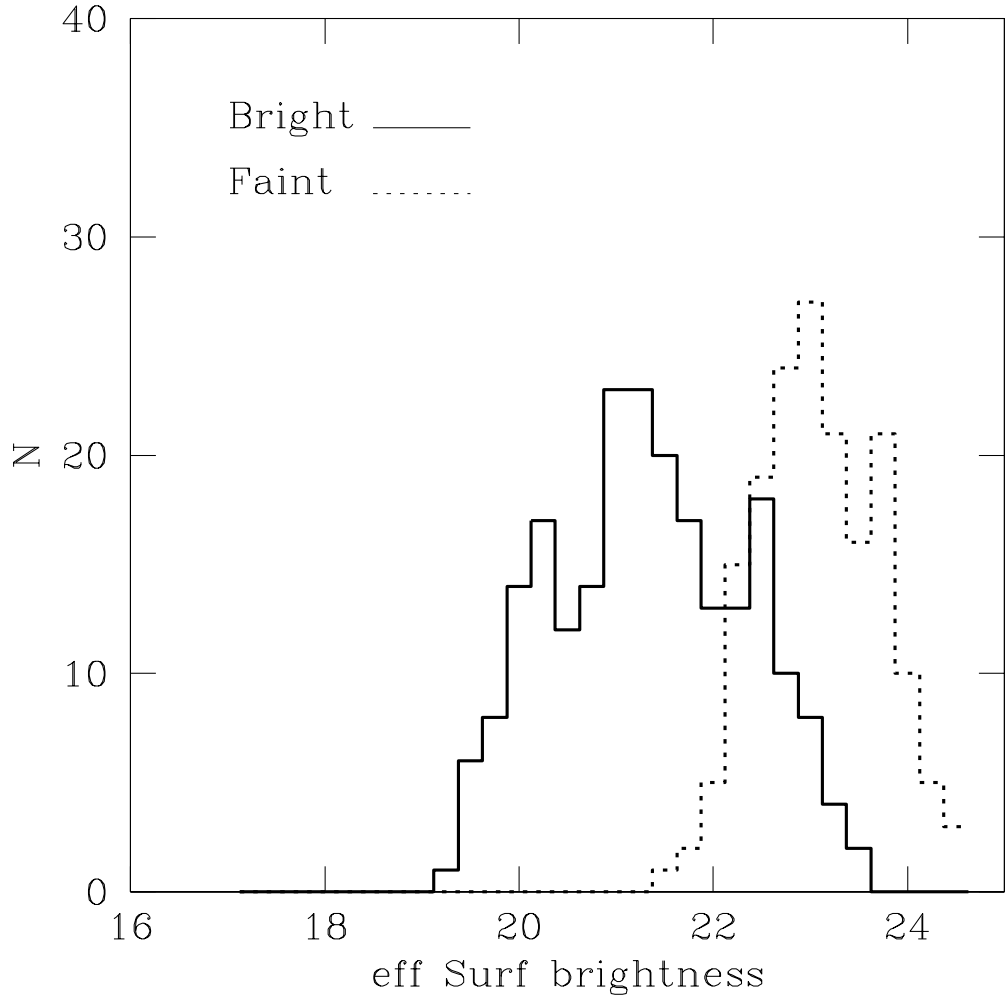}
\caption{Distribution of the R-band effective surface brightness
for the bright sample (solid line), dominated by giant galaxies, compared to
the distribution for the faint sample (dotted line), dominated by the dwarfs.
\label{fig6}}
\end{figure}
  
\subsection{Spectroscopic Observations and Data Reduction}

The spectroscopic observations for this sample were obtained with 
the WYFFOS multi-fibre spectrograph on the
William Herschel Telescope (WHT) on La Palma during April 18-23, 1998.
WYFFOS has $\sim$ 110 fibres of 2.7 arcsec diameter coupled to
a bench-mounted spectrograph of Baranne design (e.g. Bridges 1998).  
We allocated typically 60-70 objects to fibers in a given configuration, and
on average, 15-20 sky fibers. The
600B grating was used, giving a dispersion of $\sim$ 3 \AA/pixel, a total
spectral coverage of $\sim$ 3000 \AA~ with a TEK 1024$^2$ CCD,
and a resolution
varying between 6--9 \AA~ FWHM, depending on location on the CCD.  
Our spectra were centered on 5100 \AA, thus covering
many interesting spectral features ranging from Ca H\&K in the blue
to NaD in the red. 

The galaxies were divided into different configurations, depending on their
luminosities. 
A total of 10 configurations were executed for the Coma1 and Coma3 fields, 
with total exposure times of 3 hours for the
faint sample and 1 hour for the bright sample.  Argon lamps for 
wavelength calibration and offset sky exposures for fibre throughput
calibration were also obtained.  We also observed
spectroscopic flux standards, Lick standards, and radial velocity
standards during twilight.  The reduction of these multi-fibre spectra
was performed using the dedicated {\tt wyfred} package.  

Twilight sky flats or combined object
frames were used to define the apertures and trace the spectra
on the CCD.  
The median offset sky exposures were then used to calculate
the throughput for each fibre, and to normalize all of the sky fibres.
The arc spectra were extracted and matched with arc lines to
determine the dispersion solution. The fibre median {\it rms}  
ranged between 0.15--0.2 \AA~ ($\sim$ 1/20 of a pixel).
Finally, the object spectra were extracted, normalized, and wavelength 
calibrated.
The sky subtraction was performed using a master sky spectrum, 
constructed by combining the   
normalized sky spectra.   
The sky subtraction accuracy was
quite good, ranging between 1--3\% (defined as the {\it rms} of the normalized
sky fibres about the master sky spectrum). 
Sample spectra of galaxies with different luminosities in the bright and
faint samples are shown in Fig.~7.

\begin{figure}
\plotone{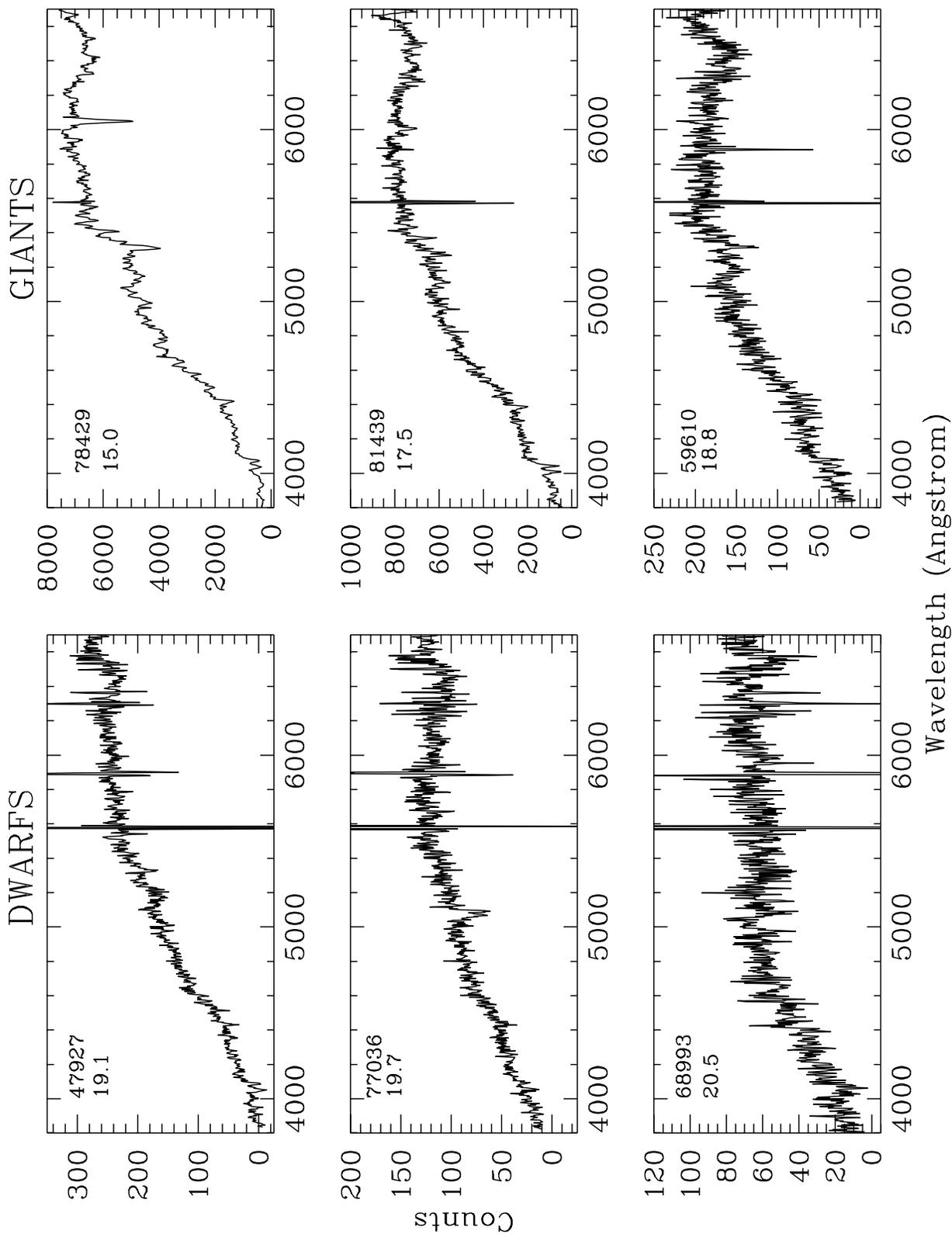}
\caption{Sample spectra of galaxies in the faint   
(left hand side) and bright
(right hand side) samples. The order is from high S/N to low S/N from top 
to bottom. The 
object IDs are shown on each panel.\label{fig7}}
\end{figure}

The spectra were flux calibrated using standard FIGARO 
procedures on the spectra of the 
flux standards, obtained on 5 of the 6 nights. Each flux standard was 
observed 
through one fibre only. As a result, fibre to fibre variations in throughput 
as a function of wavelength will cause some small systematic errors 
in the final
flux calibrated spectrum. Comparison between the flux calibration curves 
derived
from the flux standards obtained on the 5 nights shows errors of up to 
20\% in the {\em relative} flux at the two ends of the spectrum (4000 
and 6500 {\AA}). This could be due to a relative difference in 
fiber throughput as a function of wavelength.  
Our spectral indices are of course
determined over much shorter regions of the spectrum, with continuum passbands
on either side of the feature, and hence are not seriously affected by this. 

\subsection{Velocity Measurements}

Velocities are needed in this study to determine membership in
the Coma cluster (for the faint sample), and to measure the  
{\it rest-frame} spectral indices (for all the objects).
For galaxies with emission lines, velocities were determined from
the average of all measureable emission lines.  For those objects 
without emission lines,
velocities were obtained by cross-correlating 
against the radial velocity template stars hd65934, hd86801, and 
hd90861.  We used the IRAF {\tt fxcor} Fourier cross-correlation
package, masking out regions of
the spectra affected by night-sky lines.  The final velocity was
taken as that of the template with the highest cross-correlation
(Tonry \& Davis 1977) coefficient. Those with cross-correlation 
coefficients $<$ 2 were
not used.  For objects with both emission and absorption lines, 
we checked to ensure that there were no significant shifts between
emission and absorption line velocities.

As an external check on the accuracy of our velocities, the redshifts
for Coma cluster galaxies in the present survey are compared
with redshifts for the same galaxies by Colless \& Dunn (1996). 
There are a total of 147 (Coma1) and 44 (Coma3) galaxies in common
between the two surveys.
The difference between the velocities estimated in the two studies
is plotted in Figure 8 as a function of R-band magnitude 
and $B-R$ color. This shows a mean difference of
$<V_{this\ study} - V_{CD}> =$ 22 km/sec (Coma1)
and 30 km/sec (Coma3) with no dependence on galaxy magnitude
or color.

\begin{figure}
\plotone{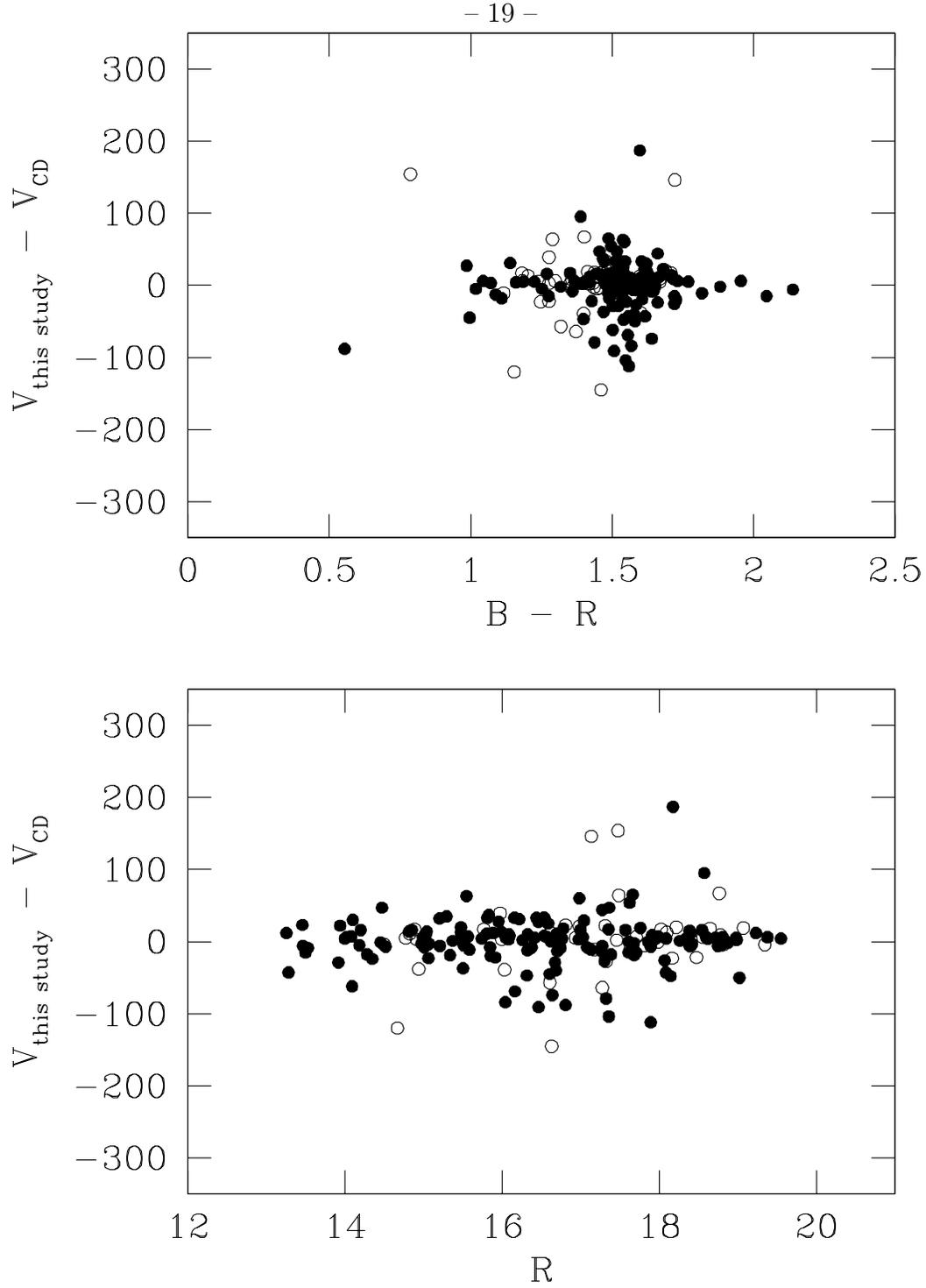}
\caption{Differences between the velocities estimated
in this study and those in Colless \& Dunn (1996)- ($V_{CD}$) is presented
as a function of R-band magnitude and $B-R$ color for Coma1 (solid
circles) and Coma3 (open circles).\label{fig8}}
\end{figure}

Assuming a Gaussian distribution of velocity differences, 
an {\it rms} of 101 km/sec is found. Therefore, an error of 100 km/sec is 
taken for the velocities of the Coma galaxies in this study, 
consistent with
that expected for our spectral resolution and S/N.  This error
corresponds to the velocities estimated for galaxies in the bright sample, 
as the 
comparison in Figure 8 is mostly dominated by these objects.
The faint sample has a relatively lower S/N, and the velocity
uncertainties will likely be higher.   
Measurement of the spectral line indices  and their associated errors
will be discussed in detail in paper III. 

The redshift distributions for {\it all} the galaxies with spectroscopic data
from this study are presented in Figure 9. This consists of the bright
and faint samples (Figures 9a and 9b) and the sub-sample selected with
no $B-R$ color criteria (Figure 9c).  
Taking a mean velocity of ~7000 km/sec and a dispersion of ~1000 km/sec
for galaxies in the Coma cluster (Colless \& Dunn 1996), 
we define a 3$\sigma$ cut corresponding
to a velocity range of 4000 $<$ V $<$ 10,000 km/sec for galaxies in order
to be included as members of the Coma cluster.
This is clear in the peak in the redshift distribution in Figure 9. 
All the galaxies which are identified as 
members of the Coma cluster in Figure 9c (where no color constraint is used)
have $B-R < 2$. 
Considering the sample selected where no color constraint was applied, as
shown in Figure 4 and Table 2, we find that 40\% of galaxies in Coma1
are confirmed cluster members, compared to only 8\% in Coma3.  
This shows the effectiveness of $B-R$ colors 
in selecting potential cluster members 
in the cluster outskirts (ie. Coma3) where contamination
by field galaxies increases. The
total number of galaxies in each of these samples and the numbers 
of Coma cluster members are listed in Table 2. 
The velocity distributions for confirmed members of the Coma cluster
are presented in Figure 10 for the same samples as in Figure 9. While
in all three samples there is a peak at 7000 km/sec, the
velocity dispersion in the Coma3 field (around the cD galaxy NGC4839)
is significantly smaller than that in Coma1. This is consistent with
the presence of a bound population of objects, forming the NGC 4839 group, 
which may be infalling towards the center of the Coma. 
The implications of this for the internal dynamics of the Coma cluster
will be studied in a future paper in this series.

\begin{figure}
\plotone{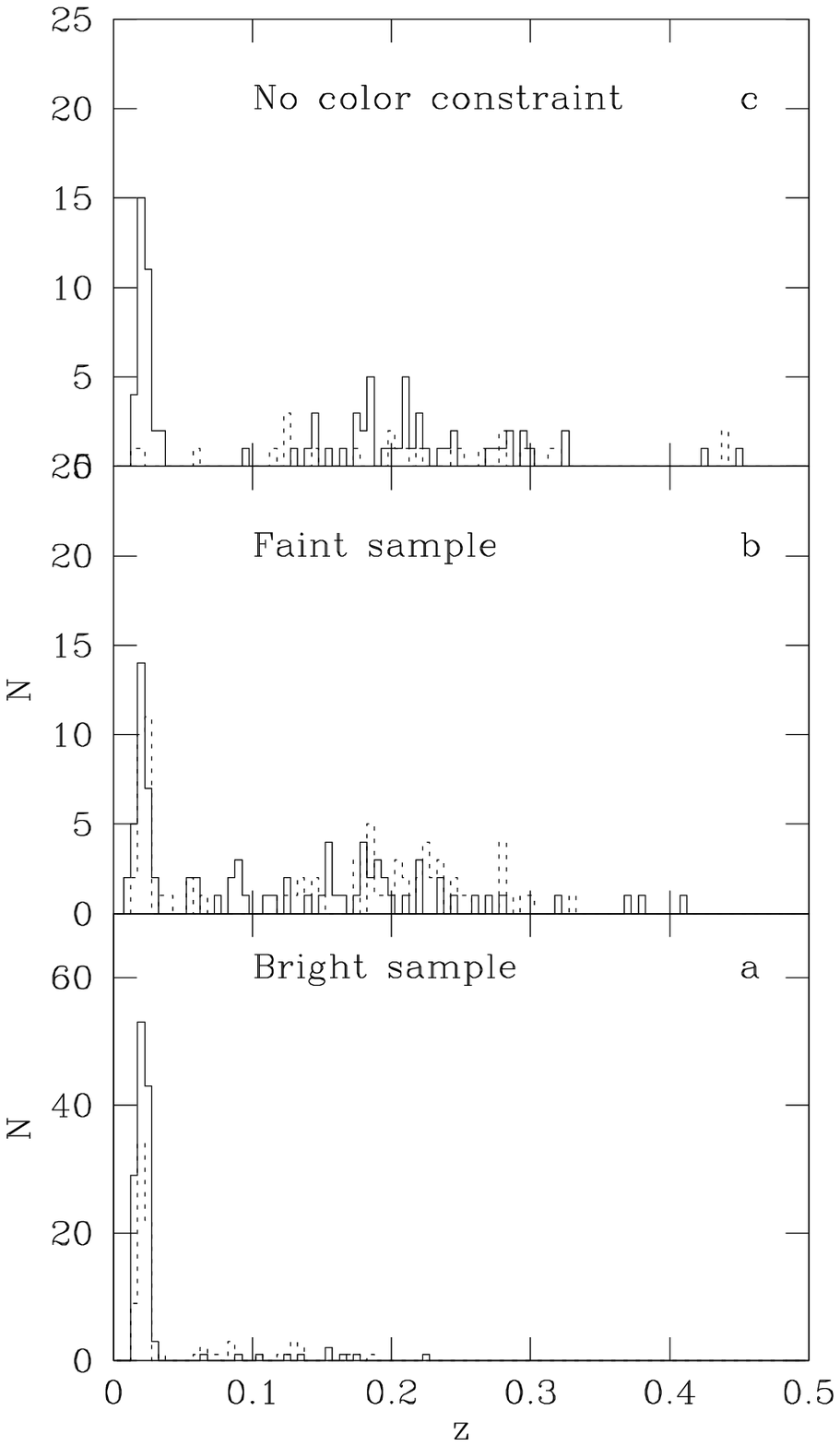}
\caption{Redshift distributions for all the galaxies observed 
in the spectroscopic survey. The comparison is given for the three different
samples with different selection criteria in this study.  Objects in the Coma1   
(solid line) and Coma3 (dotted line) fields are shown.\label{fig9}}
\end{figure}

\begin{table*}
\pagestyle{empty}
\caption{Total number of galaxies in the spectroscopic sample and the Coma cluster
members}
\begin{tabular}{lcc}
                    &            &     \\
                    &      Coma1  &     Coma3 \\ 
                    &             &        \\
Bright Sample       &             &   \\
                    &             &   \\
Total sample       &      138     &      85 \\
Coma members        &      128     &      65 \\
                    &             &   \\
Faint Sample        &             &   \\
                    &             &   \\
Total sample      &      83      &     79 \\
Coma members        &      28      &     23 \\
                    &             &   \\
No color         &             &  \\
constraint                 &              & \\  
                    &             &   \\
Total sample      &      81      &     24 \\
Coma members           &   33        &  2 \\

\end{tabular}
\end{table*}

The handful of galaxies in the bright sample which have higher redshifts 
than that of the Coma cluster
are either from the sample for which no redshifts were available (included
to allocate the spare fibers), or are those with discrepant redshifts as 
compared to the Colless \& Dunn compilation, from which the spectroscopic 
candidates were selected. The R-band magnitude and $B-R$ color distributions 
for the spectroscopically confirmed members of the cluster (both Coma1
and Coma3 fields) are presented for the bright sample (Figure 1), faint
sample (Figure 3) and the sub-sample for which no color restriction
was applied (Figure 4). 

\begin{figure}
\plotone{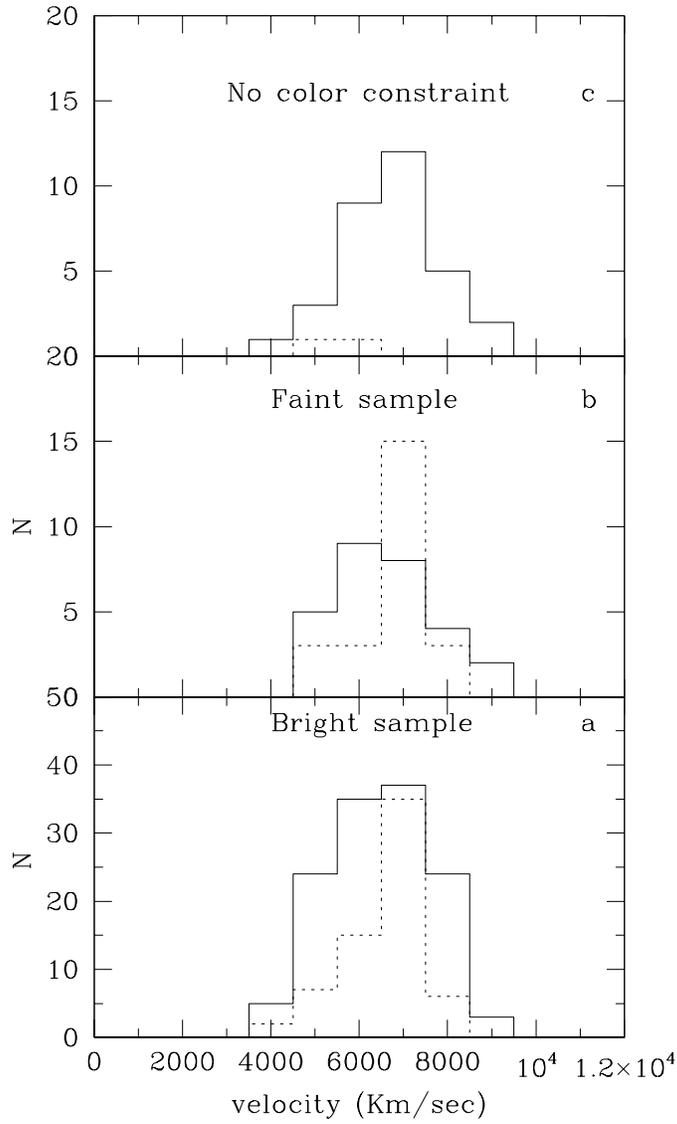}
\caption{The same as Figure 9 for confirmed members of the Coma cluster.
The distributions of galaxies in both the Coma1   
(solid line) and Coma3 (dotted line) fields are shown.\label{fig10}}
\end{figure}

The photometric and spectroscopic data for galaxies in the spectroscopic 
survey are presented in Table 3. 
The table includes the galaxy ID number (column 1), RA and Dec
in J2000 (columns 2 and 3), distance from the cluster center [taken to be at
$\alpha =12h\ 59m\ 42.8s$ and $\delta =  27^\circ\ 58'\ 14''$ (J2000)]
in degrees (column 4), R-band magnitudes and $B-R$ colors measured
over a circular aperture with a diameter of 3 Kron radii 
(columns 5 and 6), the Kron radius in arcsec (column 7), 
R-band mean and effective surface brightness (columns 8 and 9),
redshift (column 10), and cluster membership (column 11). 
Measurement of the spectral line strengths for these galaxies are 
discussed in paper III, with the line strengths given in a future paper in
this series.

\section{The completeness functions for the spectroscopic survey}

Knowledge of the completeness function is needed for the future analysis 
of the present spectroscopic sample. Since two different selection criteria
were used to conduct the spectroscopic survey, 
different completeness functions are needed
for each of these groups.

For the bright sample, the completeness function in each luminosity interval
is defined 
as the ratio of
the number of galaxies which are {\it spectroscopically} confirmed
members of the Coma cluster to the {\it expected} number of galaxies in
Coma. The expected number of galaxies in the Coma cluster
is estimated as the difference between the total number of galaxies
in a given magnitude interval  
in the photometric survey and the number of galaxies in the same
magnitude interval which are detected in the control field, SA57. The
control field is adopted to be close to the Coma cluster and is observed
in exactly the same way and to the same depth as the Coma fields.   
The completeness function for this sample in the combined Coma1
and Coma3 fields is presented in Figure 11.  

For the faint sample, the completeness function in a given 
magnitude interval is defined as the ratio of the number of galaxies 
measured spectroscopically to the total number of galaxies in the photometric 
survey with 1 $<$ B$-$R $<$ 2. 
%Changes in this ratio with luminosity gives the completeness   
%function for this survey. 
The spectroscopic completeness function for the 
faint sample in the
combined Coma1 and Coma3 fields is also presented in Figure 11. It is
clear that the bright sample is $65\%$ complete to R=17 mag. 
with the completeness
decreasing rapidly at fainter magnitudes, while the faint sample 
follows a monotonic
completeness function. The faint 
sample is 60\% complete
at 18 mag, becoming increasingly incomplete towards fainter 
($R\sim 19.5$) magnitudes. 

\begin{figure}
\plotone{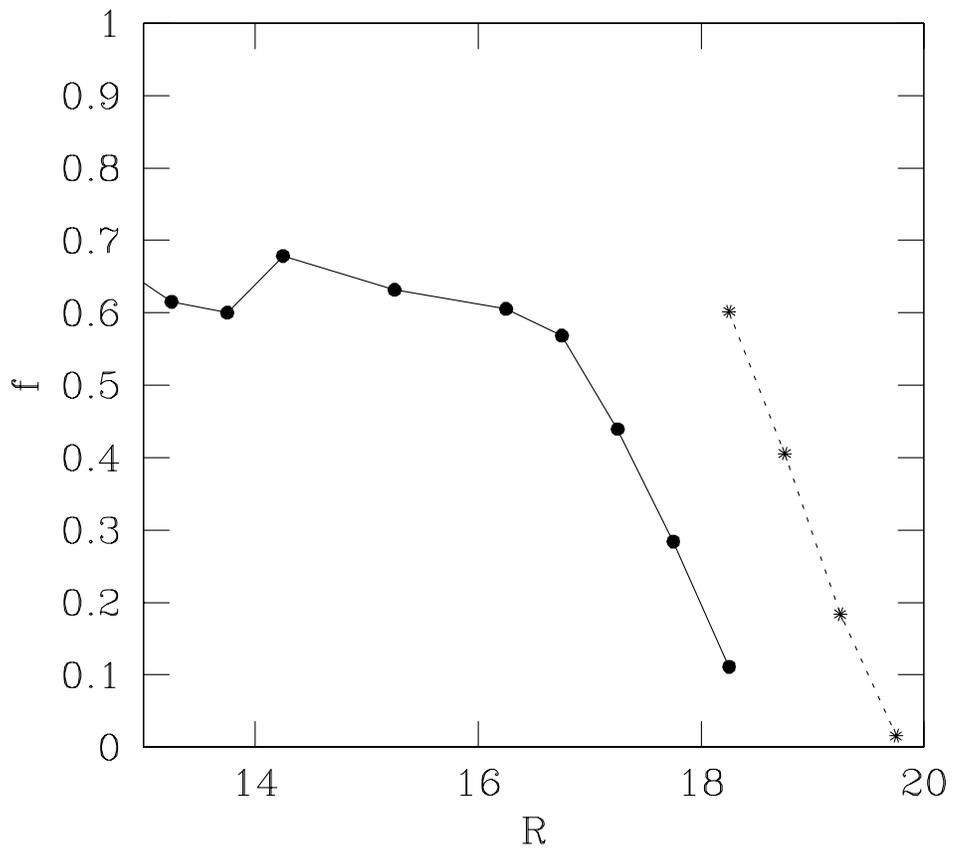}
\caption{The spectroscopic completeness function for 
the bright (solid line) and faint (dotted
line) samples.\label{fig9}}
\end{figure}

\section{Summary}

Medium resolution spectroscopy (8-9 \AA) was carried out for a sample of 490
galaxies of different morphological types at the core (302 galaxies) 
and outskirts (188 galaxies) of the Coma cluster. The galaxies cover a range
in magnitude of $ 12 < R < 20 $, corresponding to $-23 < M_R < -15$ 
(H$_0$=65 km/sec/Mpc). The candidates for spectroscopic observations were 
selected from our wide-angle photometric survey of the Coma cluster
(paper I). Three sub-samples, with different selection criteria, 
were used for this purpose, consisting of;
a) the `bright sample' with $ R < 18$, consisting of galaxies 
(drawn from a magnitude limited sample) which are
spectroscopically confirmed members of the Coma cluster. For this sample, 
the medium resolution spectroscopy allows measurement of the spectral lines; 
b) the `faint sample' satisfying the criteria $17.5 < R < 20$ and
$1 < B-R < 2$. For these galaxies, we measure both the redshifts and 
spectral line indices;
c) A random sample of galaxies with no available redshifts,  
where the $B-R$ color constraint is relaxed. This is to examine if 
the selection criteria for the `faint sample' would exclude cluster members. 

The total number of galaxies identified as spectroscopically confirmed members
of the Coma cluster is 189 (at the core) and 90 (at the outskirts).
Only two galaxies (in Coma1) are found with
$B-R > 2$ which are members of the cluster. 
An analysis of the completeness function for the spectroscopic 
survey shows that
the `bright sample' is $65\%$ complete at $R < 17$, becoming increasingly
incomplete towards fainter magnitudes, while the `faint sample' shows
a monotonically decreasing completeness function in the range
$18 < R < 20$.

\clearpage

\begin{deluxetable}{lllllllllll} 
\tablewidth{0pt}
\tablecaption{Table 3. List of the galaxies in the spectroscopic survey in
Coma1(C1) and Coma3 (C3). The complete table will be presented in the 
ApJS paper.}
\tablehead{\colhead{ID } &  
\colhead{RA (J2000)}&\colhead{Dec (J2000)}&\colhead{$r$}&
\colhead{R}&\colhead{B-R}&\colhead{$r_{Kron}$}&\colhead{$\mu_{mean}$}&
\colhead{$\mu_{eff}$}&
\colhead{z}&
\colhead{comment}}   
\startdata
        &                 &                 &        &        &       &        &  
&       &  \\
  Coma1   &                 &                 &        &        &       &        &   
&       &  \\  
   1144 & 13 \  0 \  4.95 & 27 \ 33 \ 10.48 &  0.426 & 18.152 & 2.503 &  1.496 & 22.72  & 20.53 & 0.3010   \\
   1230 & 13 \  0 \ 27.50 & 27 \ 33 \ 16.09 &  0.448 & 19.550 & 1.970 &  1.326 & 23.26  & 22.06 & 0.5244   \\
   2149 & 13 \  0 \ 25.16 & 27 \ 33 \  7.89 &  0.447 & 16.080 & 1.536 &  3.743 & 22.45  & 20.51 & 0.0241 C1    \\
   2179 & 13 \  0 \ 24.86 & 27 \ 33 \ 42.86 &  0.437 & 19.332 & 0.979 &  1.976 & 23.62  & 22.64 & 0.0999   \\
   2277 & 12 \ 58 \ 41.26 & 27 \ 33 \ 34.90 &  0.469 & 18.457 & 1.379 &  2.127 & 23.10  & 21.84 & 0.1854   \\
   2623 & 12 \ 59 \  6.38 & 27 \ 33 \ 38.94 &  0.431 & 17.860 & 1.483 &  3.585 & 23.66  & 22.43 & 0.0265 C1    \\
   2938 & 12 \ 58 \ 39.46 & 27 \ 34 \  0.69 &  0.466 & 19.045 & 2.068 &  1.539 & 23.27  & 22.05 & 0.2860   \\
   3061 & 13 \  0 \  9.69 & 27 \ 33 \ 59.10 &  0.416 & 18.151 & 1.450 &  1.863 & 22.80  & 21.31 & 0.1653   \\
   3368 & 12 \ 59 \ 30.74 & 27 \ 34 \ 15.93 &  0.402 & 19.920 & 1.165 &  1.282 & 23.49  & 22.42 & 0.1884   \\
    .   &  . \ . \ . \    & . \ . \ . \     &    .   &   .    &   .   &    .   &   .    &   .   &   .      \\
    .   &  . \ . \ . \    & . \ . \ . \     &    .   &   .    &   .   &    .   &   .    &   .   &   .      \\
    .   &  . \ . \ . \    & . \ . \ . \     &    .   &   .    &   .   &    .   &   .    &   .   &   .      \\
        &                 &                 &        &        &       &        &

&       &  \\
  Coma3   &                 &                 &        &        &       &
 & 
&       &  \\
&                 &                 &        &        &       &        &
&       &  \\
   3089 & 12 \ 58 \ 24.41 & 26 \ 43 \ 46.44 &  1.274 & 20.284 & 2.006 &  1.199 & 23.57  & 22.66 & 0.4410   \\
   3137 & 12 \ 56 \ 27.16 & 26 \ 43 \ 37.91 &  1.439 & 18.493 & 1.626 &  1.530 & 22.78  & 21.17 & 0.3015   \\
   3160 & 12 \ 58 \ 22.57 & 26 \ 43 \ 42.06 &  1.277 & 19.375 & 2.845 &  1.369 & 23.37  & 21.61 & 0.3181   \\
   3531 & 12 \ 57 \ 47.14 & 26 \ 43 \ 43.43 &  1.314 & 18.367 & 1.256 &  1.406 & 22.65  & 20.85 & 0.1772   \\
   3611 & 12 \ 56 \ 19.15 & 26 \ 43 \ 32.34 &  1.455 & 16.864 & 1.184 &  3.353 & 22.72  & 21.21 & 0.0790   \\
   6170 & 12 \ 58 \ 19.14 & 26 \ 44 \  4.04 &  1.274 & 17.584 & 1.200 &  3.899 & 22.88  & 21.74 & 0.0708   \\
   8304 & 12 \ 56 \ 37.00 & 26 \ 44 \ 32.90 &  1.407 & 17.999 & 1.577 &  3.096 & 23.59  & 22.67 & 0.0203 C3    \\
   9487 & 12 \ 56 \ 43.23 & 26 \ 44 \ 31.57 &  1.397 & 15.863 & 1.495 &  3.268 & 22.32  & 19.60 & 0.0233 C3    \\
  12996 & 12 \ 57 \ 50.07 & 26 \ 45 \ 50.77 &  1.277 & 18.468 & 1.582 &  1.482 & 22.97  & 20.57 & 0.1902   \\
    .   &  . \ . \ . \    & . \ . \ . \     &    .   &   .    &   .   &    .   &   .    &   .   &   .      \\
    .   &  . \ . \ . \    & . \ . \ . \     &    .   &   .    &   .   &    .   &   .    &   .   &   .      \\
    .   &  . \ . \ . \    & . \ . \ . \     &    .   &   .    &   .   &    .   &   .    &   .   &   .      \\
\enddata
\end{deluxetable}

\clearpage

\acknowledgments
We are grateful to Matthew Colless and Richard S. Ellis for helpful 
discussions during the course of this work. The spectroscopic data presented 
in this paper are based on observations made with the William Herschel
Telescope operated on the island of La Palma by the Isaac Newton Group in
the Spanish Observatorio del Roque de los Muchachos of the Instituto 
de Astrofisica de Canarias.

\end{document}